# Device-scale perpendicular alignment of colloidal nanorods

*Jessy L Baker[1], Asaph Widmer-Cooper[2], Michael F Toney[3], Phillip L Geissler[2], A Paul Alivisatos[2]\**

Departments of Mechanical Engineering[1] and Chemistry[2], University of California, Berkeley, Berkeley California 94720, and Stanford Synchrotron Radiation Lightsource[3], Palo Alto, California 94025

Corresponding author: alivis@berkeley.edu.

Keywords: grazing-incidence x-ray diffraction, nanorod, vertical alignment, self-assembly

The self-assembly of nanocrystals enables new classes of materials whose properties are controlled by the periodicities of the assembly, as well as by the size, shape and composition of the nanocrystals. While self-assembly of spherical nanoparticles has advanced significantly in the last decade, assembly of rod-shaped nanocrystals has seen limited progress due to the requirement of orientational order. Here, the parameters critically relevant to self-assembly are systematically quantified using a combination of diffraction and theoretical modeling; these highlight the importance of kinetics on orientational order. Through drying-mediated self-assembly we achieve unprecedented control over orientational order (up to 96% vertically oriented rods on 1cm$^2$ areas) on a wide range of substrates (ITO, PEDOT:PSS, $Si_3N_4$). This opens new avenues for nanocrystal-based devices competitive with thin film devices, as problems of granularity can be tackled through crystallographic orientational control over macroscopic areas.

Colloidal nanocrystals offer a potential route to realizing low-cost solution-processed electronic devices. In particular, with their size-tunable properties, single-crystallinity, and inexpensive synthesis, semiconductor nanoparticles could enable improved optoelectronic devices. To date, however, the performance of nanoparticle devices has been limited, in large part, by the number of interfaces that charge carriers encounter before they can be collected; each interface presents an opportunity for recombination and subsequent charge loss or relaxation. An ideal film would appear as a single-crystal to charge carriers, but would maintain solution-processability and therefore low manufacturing cost. This has been attempted with nanowire arrays[1-3], which can have single-crystal properties in the through-film direction (necessary for optoelectronics), but such arrays have met with only partial device success due to the low nanowire packing density and difficulty of incorporating semiconductor material between wires. Using colloidal nanocrystals, researchers have explored carrier mobility and device performance in the context of low-resistivity percolation networks. Improving the percolation network (low-resistivity pathway for charge carriers) has steadily improved performance as devices have evolved from using spheres to randomly-oriented rods to hyperbranched particles[4,5]. An ideal device geometry would consist of a monolayer of vertically oriented rods (perpendicular to substrate) spanning the full thickness of the film. While vertical nanowires have been formed on substrates using batch processing[6-9], the aim of this work was to achieve a perpendicular morphology in one step using solution-processable nanocrystals with no pre-patterning requirements or substrate restrictions.

Several reports have presented the perpendicular alignment of nanorods on a small scale ($\mu m^2 s$)[10-16], but larger areas necessary for devices have not been reported quantitatively, nor is there a firm understanding of the underlying physical principles directing self-assembly, which would facilitate rational design. A key component missing from the literature until recently is large-scale quantification of nanorod orientation[17]. Here, we report a slow-drying method for self-assembly of colloidal semiconductor nanorods resulting in their device-scale perpendicular alignment. We explore the self-assembly parameter space, and discuss the important role kinetics plays in alignment. This assembly method is general to a variety of substrates and can be expanded to the square centimeter scale.

In this work, nanorod films were produced with both short- and long-range order by controlling the evaporation of a solution of cadmium sulfide (CdS) nanorods (Figure 1). CdS nanorods (4nm in diameter, tunable 30-100nm in length, with surfaces passivated by octadecylphosphonic acid) were used in these studies due to the well-established synthesis[18] of single-crystal samples monodisperse in size and shape. Before alignment, the nanorods were stored air-free, and rod concentrations were measured by UV-vis absorption calibrated with inductively coupled plasma mass spectrometry. With a slow evaporation rate (<1mm/min meniscus speed across substrate), an elevated temperature (55°C), and an appropriate substrate (for example, silicon nitride), nanorods oriented vertically. (See Supporting Information for further details.) To better understand the key parameters controlling self-assembly and thus aid rational device design, electron microscopy images were paired with a diffraction measurement technique representative of full-film morphology.

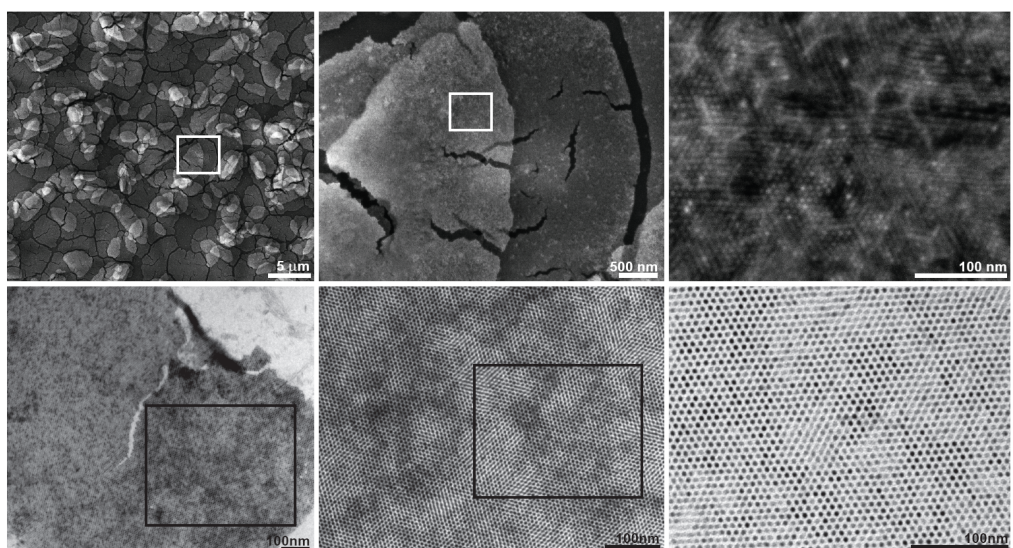

**Figure 1.** Electron microscopy images of large vertically oriented rod domains with progressive zoom from left to right. Top/bottom row: SEM/TEM images. Regions of higher contrast correspond to multiple monolayers. At this scale, assembly is difficult to judge quantitatively with electron microscopy.

Assembly of quantum *dots* can be assessed with a small-angle x-ray diffraction measurement (which probes the particle-to-particle periodicity, rather than lattice periodicity in individual nanoparticles) because particle rotational orientation is unimportant; only positional information is needed[19]. For information about nanocrystal crystallographic orientation, and for facile data interpretation, wide-angle diffraction is an alternative technique. In some cases, these methods are complementary.

Grazing-incidence wide-angle x-ray diffraction (GIXD) enables morphological quantification of assembled nanoparticle thin-films by leveraging the anisotropic lattice common to many semiconductor nanocrystals (Figure 2). Here, in the case of cadmium sulfide (wurtzite lattice), the long axis of the nanorod corresponds to the c-axis ( (002) direction) in the rod's lattice. Therefore, knowledge of the *lattice* orientation translates to knowledge of the *rod* orientation. While any diffraction peak could be used to monitor rod orientation, (002) was chosen because of the sharp signal afforded by its orientation along the long axis of the rod. (002) diffraction from rods oriented parallel to the substrate falls at the horizon of the Bragg ring in the diffraction pattern (Figure 2 c and d) and corresponds to rod orientation angle $\omega=\pm90°$. Diffraction from rods oriented perpendicular to the substrate falls at the top of the Bragg ring, where $\omega$ approaches $0°$. Radial cross-sections of the diffraction pattern taken at different angles show the degree of anisotropy in nanorod orientation (Figure 2e and f). Integrating diffraction intensity as a function of angle $\omega$ along the circumference of the (002) Bragg ring gives us an orientation distribution function (Figure 2g and h).[20]

By using a large X-ray beam size and rastering the beam across the substrate, we obtain an orientation distribution function (ODF) representative of the entire film. By integrating the normalized and intensity-corrected ODF[21] within a range of angles of interest, one can say with precision how many rods are oriented within this range. Here, vertical is considered to be $\pm20°$, as this amount of tilt corresponds to a small change in film thickness of only 6%.

Strong texture is observed (intensity variation around Bragg ring) in the diffraction patterns from films of vertically oriented nanorods (Figure 2d), corresponding to a narrow ODF indicating vertical nanorod alignment. The width of the orientation peaks may be due to contraction of the ligand shell upon final solvent drying, causing rods to lean slightly. This could be understood as a cumulative effect of the shrinkage of ligands on many rods (approximately 0.4nm per rod), creating (over the area of tens of rods) enough space for either cracks or rod inclination (see cracks and tilting rods in Figures 1 and S3, respectively).

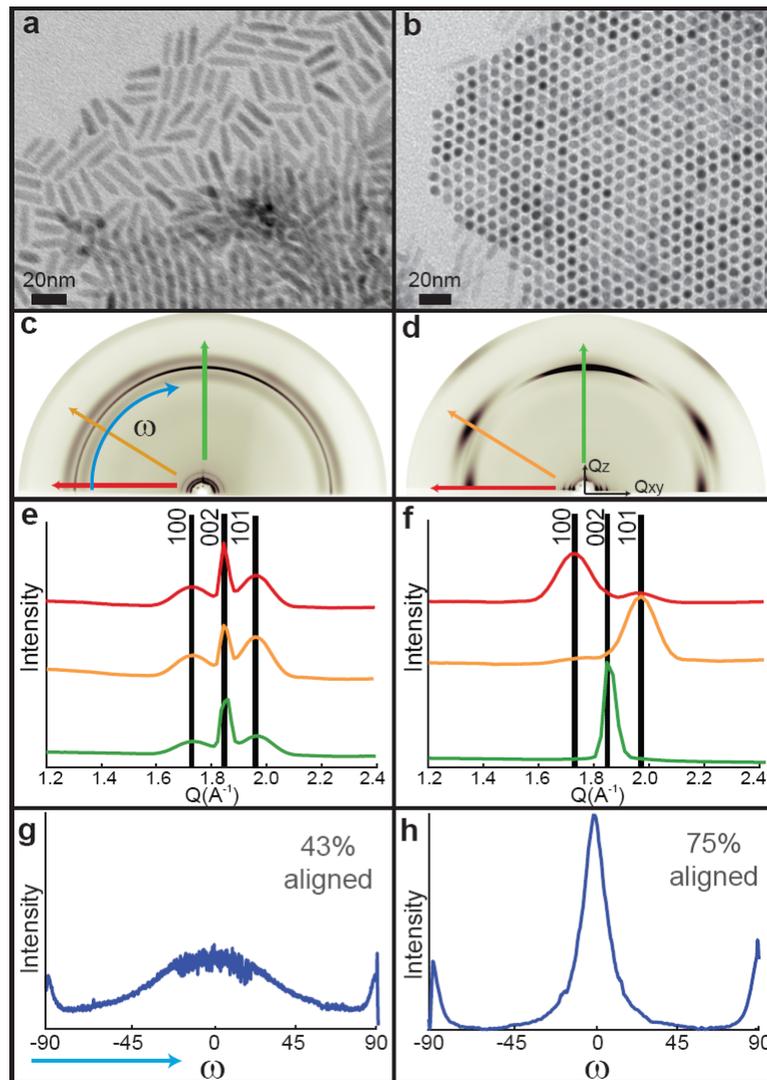

**Figure 2.** X-ray diffraction is a necessary complement to electron microscopy for film morphology assessment. (a) and (b) TEM images of near-isotropic and vertically aligned nanorods. The c-axis of the anisotropic wurtzite lattice is oriented along the long axis of nanorods, enabling detection of rod orientation with wide-angle diffraction. (c) and (d) Diffraction patterns corresponding to morphologies (a) and (b). Intensity collected near the horizon of the (002) (middle) Bragg ring is diffraction from horizontally oriented rods, while intensity near the top is from vertically oriented rods. Low-Q intensity is from small-angle diffraction events (superlattice). (e) and (f) Radial cross-sections of diffraction patterns shown in (c) and (d) indicate an angle dependence only in the case of an oriented film. (g) and (h) Orientation distribution obtained by plotting the intensity of rocking and grazing data along the (002) Bragg rings in (c) and (d) as a function of ω[21,22]. From these plots, we can determine the percentage of rods vertically oriented (within ± 20° of the normal to substrate): 43% and 75% are vertically aligned in (g) and (h). The tails around ω=±90° in Figure 2h indicate that some of the rods lie horizontally.

To quantitatively explore the dependence of rod alignment on key assembly parameters (temperature, aspect ratio, and van der Waals (vdW) interaction with the substrate (Figure 3)), we used the GIXD analysis described in Figure 2 together with theoretical calculations of interaction energies and diffusion constants. Elucidation of the contributions from entropy, enthalpy, and kinetics gave us insight into the physics governing rod self-assembly. The standard experimental conditions for data plotted in Figure 3 were: 55° C nanocrystal solution temperature, rod aspect ratio 6.4, rod diameter 4nm, polydispersity σ=12%, tetrachloroethylene as solvent, $Si_3N_4$ substrates, and evaporation rates resulting in ~1mm/min meniscus speeds across the substrate. The total number of nanorods in solution was chosen to provide at least one full monolayer of vertically aligned rods on the final substrate. To avoid aggregation, experiments were performed in the presence of excess surfactant (see Supporting Material for additional experimental details).

Heating the nanorod solution 10-40°C above room temperature (while keeping the evaporation rate constant) favored perpendicular rod alignment (Figure 3a), consistent with previous qualitative reports[10,14]. This may point to a kinetic effect and/or competition between entropic and energetic driving forces. The rotational and translational diffusion constants ($D^i$) for rod-shaped particles in solution[23] depend on temperature as $D^i \propto T/\eta_s$, where $\eta_s$ is the shear viscosity. Since $\eta_s$ decreases upon heating, self-assembly will become faster at higher temperature, effectively giving the system more time to lower its free energy. In this case, the increase in alignment upon heating would indicate that the perpendicularly aligned state is thermodynamically stable. The trend in Figure 3a could also indicate that there are entropic effects driving perpendicular alignment that are opposed by enthalpic interactions. It is well known that hard rod-shaped particles will spontaneously undergo orientational ordering when their containing volume shrinks (e.g. due to solvent evaporation), because the gain in positional entropy on alignment (associated with reducing the total excluded volume) more than compensates for the loss in rotational entropy[24]. Li *et al.* have observed similar liquid-crystalline behavior for CdSe nanorods[25], which have analogous physical properties to the CdS nanorods used in the present study. In Li's work, the isotropic-nematic phase boundary was found to be weakly temperature dependent over the range considered (some of this data is plotted in Figure 3a for comparison), indicating that the entropically driven ordering may be opposed by enthalpic interactions that become less important at higher temperature.

Similar to previous reports, we found that low aspect ratio (AR) rods aligned more readily than rods with high AR (Figure 3b). While we were able to align rods with ARs up to 15 (higher than any published report), we were unable to identify a set of parameters that achieved detectable vertical orientation with nanorods of AR approaching 25. This is likely a predominantly kinetic effect. The short-time (non-interacting) rotational diffusion constant $D^{rot}$ for rod-shaped particles in solution is strongly dependent upon their length $L$[23]. For the size of nanorods considered here $D^{rot} \propto \ln(L)/L^3$) and decreases by ~97% going from an AR of 5 to 25. The long-time rotational diffusion (considering rod-rod interactions) should follow a similar trend (see Supporting Material). Based on thermodynamic considerations alone, we expect little change in alignment with increasing AR. The relative energy per particle for a monolayer of CdS rods aligned perpendicular and parallel to $Si_3N_4$ changes little with AR in the range considered here (see Supporting Material). It is also known that for entropic reasons longer rods order orientationally at lower volume fraction [ref: P. Bolhuis & D. Frenkel. Numerical study of the phase behavior of rodlike colloids with attractive interactions. *J. Chem. Phys.* **107**, 1551 (1997).], which if anything, would be expected to increase alignment. The interaction energy between the rod (both the nanocrystal and the ligands) and the substrate in solution also plays a role in rod self-assembly, as shown by its effect on alignment in Figure 3c (details of the vdW energy calculations are in the Supporting Material). A strongly attractive interaction encourages rods to maximize their area of contact with the substrate by lying parallel to it. As this attraction is decreased, a stronger preference for perpendicular alignment is observed. Similar to Titov[27], we found that the substrate influenced alignment. However, our experimental results show that the alignment is not a simple function of the substrate Hamaker constant, and that in general alignment is not strongly dependent upon the choice of substrate (see also Figure 4) over the range of interactions we considered. Based on theoretical analysis, we suggest that the ligands bound to the nanorods play an important energetic role. The vdW energies calculated in solution are typically not much larger than thermal energy, primarily because the ligands prevent the crystalline cores from interacting strongly with each other and with the substrate. Note that the typical separation between rods surfaces in our experiments is 3.7 nm. This result explains why perpendicular alignment is only weakly substrate dependent. The inclusion of the ligands in our analysis also offers an explanation for the subtle dependence of alignment on the choice of substrate. If the Hamaker constant of the substrate is greater than that of the solvent then the ligands are effectively repelled by the substrate at close range. This effect leads to a non-monotonic dependence of the rod-substrate interaction energy on the substrate Hamaker constant, as shown by the curve in Figure 3c.

Our calculations also show that ligand-ligand and ligand-substrate interactions are large and dominant in the absence of solvation (100s of kT). While it is clear that self-assembly does not occur after drying is complete, fluctuations in solvent density near the liquid-vapor interface could activate these strong attractions in the late stages of assembly[28,29]. The visible film contraction and cracking that occurs at the final stage of drying is also consistent with the large increase in ligand vdW interactions that we calculate upon desolvation. Finally, our calculated interaction energies (in both vacuum and solution) suggest that perpendicular alignment should be largely independent of the choice of nanocrystal material. Indeed, reports of self-assembly demonstrate vertical alignment of CdS[12], CdSe[15], CdS-CdSe heterostructures[14], $Cu_2S$[30], $Ag_2S$[30], and Au[31] rods.

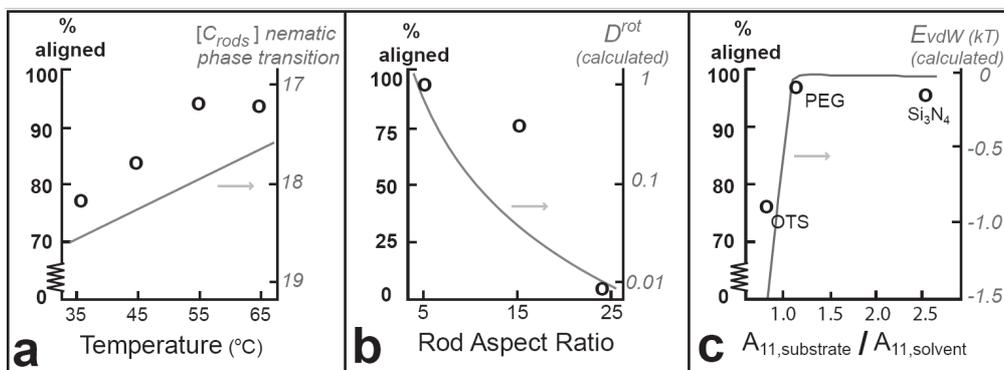

**Figure 3.** Entropy, enthalpy, and kinetics all contribute to self-assembly. a) An elevated temperature increases the percentage of vertically aligned rods (**o**'s), and decreases the required rod concentration (volume %) for a nematic (ordered) phase transition (*line*)[25]. This temperature dependence demonstrates kinetic and/or entropic contributions to self-assembly. b) Rods with higher aspect ratio are more difficult to align (**o**'s). The estimated rotational diffusion constants (normalized by $D^{rot(AR=5)}$) similarly decrease with increasing aspect ratio (*line*) suggesting that this is a kinetic effect. c) The percentage of vertically aligned rods (**o**'s) is a non-monotonic function of the substrate Hamaker constant ($A_{substrate}$) and can be explained by the rod-substrate interaction

strength. Substrates were silicon nitride TEM membranes, in some cases functionalized with self-assembling monolayers (SAMs): octadecyltrichlorosilane (OTS) & polyethylene glycol (PEG). A stronger attraction, calculated as the vdW interaction in solution for a rod parallel to the substrate (*line*), results in more horizontally oriented rods (**o**'s). Diffraction results were corroborated qualitatively with TEM images of each sample. Error bars fall within each data point. Identical nanocrystal samples and preparation procedures were used for each temperature and substrate study for consistency. Rods with AR=6.4 had a width polydispersity of 10%, and a length polydispersity of 12% (used also for temperature and substrate studies) Rods with AR=15.7 had a width polydispersity of 12.9% and a length polydispersity of 8.5%. Rods with AR=23.75 had both length and width polydispersities of 10%. [I think this additional polydispersity data would be better placed in the Supporting Material]

In summary, the strong trend counter to thermodynamic expectations that is achieved by aspect ratio variation (as compared to temperature or substrate variation) suggests that kinetics plays a dominant role in rod self-assembly. While the analysis presented here provides valuable insight into this process, a complete description will need to address the kinetic interplay among evaporation, concentration, aggregation, deposition, and flow that ultimately determines film structure and particle orientation. Drying-mediated self-assembly is inherently a complex dynamical process that involves phase transitions in both nanorod and solvent density, i.e., slowly relaxing systems far from equilibrium.

Our quantitative exploration of factors affecting alignment allowed us to optimize assembly conditions and thereby vertically align nanorods on a variety of substrates (Figure 4); low aspect-ratio rods and an elevated temperature were used to obtain better vertical nanorod alignment. Because the rod-substrate interactions are relatively weak, we were able to align rods on a wide range of substrates. Wide peaks in the ODF were observed for rough substrates (indium tin oxide (ITO) and rough silicon nitride ($Si_3N_4$); RMS roughness ~ 1.2nm and 1.7nm respectively), while sharper peaks were observed for smooth substrates ($Si_3N_4$ RMS roughness 0.5nm), demonstrating that rough surfaces do not significantly disturb rod alignment, but merely broaden the orientation distribution; rough and smooth $Si_3N_4$ substrates alike had only 1-3% of rods oriented horizontally (within 20° of the substrate plane). Interestingly, the $Si_3N_4$ substrates with a PEG SAM

(polyethylene glycol self-assembling monolayer) provided the best vertical alignment, with up to 95% of the rods vertically aligned, consistent with the vdW interaction data in Figure 3c. The variety of substrates on which rods have been demonstrated to self-assemble with a high degree of alignment points further to the generality of our method; both rods and substrates of varying compositions can participate in the drying-mediated self-assembly described here.

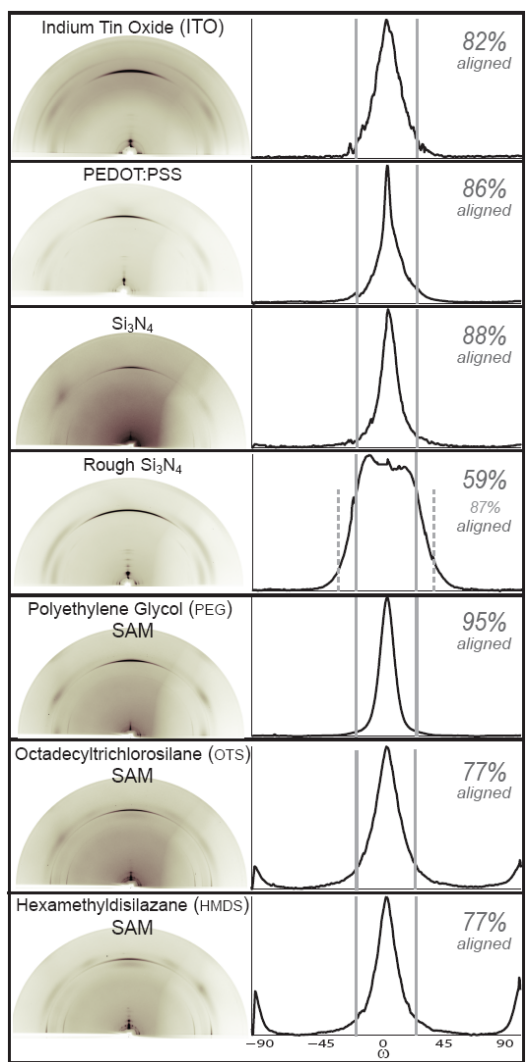

**Figure 4.** Substrate generality is shown: diffraction patterns and corresponding orientation distribution functions for each substrate. Percent vertical alignment (shown on right-hand side) is calculated by integration of the orientation distribution function between ±20°. A rough substrate results in a wider angular distribution for aligned rods. On OTS and HMDS substrates, we can see significant tails indicating horizontally oriented rods. Substrates here are ~7mm². Nanocrystal sample used was identical for each substrate, for fair comparison.

We further show scaling of our technique to larger areas, achieving 96% alignment of the nanorods on a 1cm² substrate (Figure 5). By adapting techniques from photonic crystal assembly[32] film quality was improved in terms of its uniformity in thickness and a reduction in crack size; the substrate was carefully cleaned and the evaporation rate was reduced to <1mm/min meniscus speed (across substrate). A substrate tilt angle of 5° provided a single meniscus extended across the substrate during drying, producing a uniform film of aligned nanorods across the entire area.

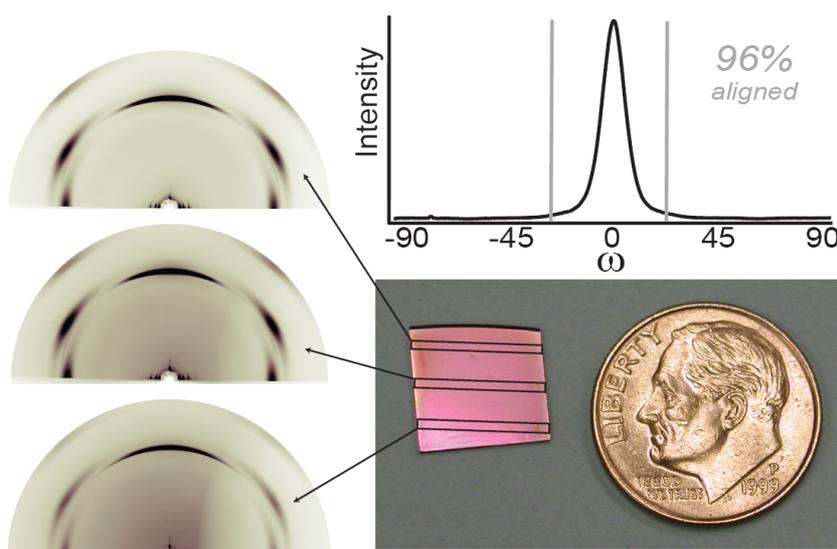

**Figure 5.** This self-assembly technique enables nanorod alignment on a device scale (1cm²). Diffraction patterns corresponding to different areas on the substrate demonstrate uniformity across the film. Integrated intensity (ODF) from these diffraction patterns shows that we are able to achieve 96% vertical alignment on this silicon nitride substrate, pictured here next to a dime.

These large-area films comprise an assembly of many smaller domains. We found that in addition to vertical orientation, the rods in each of these domains were rotationally ordered with respect to each other about their c-axis (with ±15° leeway for rotation about this axis). Selected area wide-angle electron diffraction shows this formation of a supercrystal in two dimensions (see Supporting Material Figure S5). Because wurtzite nanorods are known to have an approximately hexagonal cross section[33], their faceting might induce the rotational order observed. This result indicates that by probing an entire domain one could investigate nanoscale properties as a function of the crystal axis, currently impossible with colloidal nanorods randomly oriented either in solution or in dried films.

Further improvement in film quality will facilitate integration of these films in devices. Also, with further insight into the rod self-assembly mechanism, it may be possible to achieve alignment of rods with higher aspect ratio. This in turn would expand the applicability of this technique by making it possible to synthesize monolayer films of selectable thickness. We have characterized key self-assembly parameters, exploring the importance of enthalpy, entropy, and kinetics, and we suggest that kinetics is the dominant effect that can be modified by varying experimental parameters. This exploration led us to vertically align nanorods on many device-relevant substrates, and at the centimeter scale. Ultimately, this advance may play a critical role in the development of inexpensive, solution-processed optoelectronics with performance matching that of bulk semiconductor devices.

**Acknowledgements**. We thank Steven Volkman and Shong Yin for their help at the Stanford Synchrotron Radiation Lightsource facilities. We thank Jen Dionne for SEM images taken at Lawrence Berkeley National Labs' Molecular Foundry, and Munekazu Motoyama for AFM roughness measurements. We thank PK Jain, KM Tye, DJ Milliron, JJ Urban, JM Luther, CL Choi, JE Millstone and B Sadtler for helpful discussions and critical review of our manuscript. This work was funded by the Helios Solar Energy Research Center which is supported by the Director, Office of Science, Office of Basic Energy Sciences, Materials Sciences and Engineering Division, of the U.S. Department of Energy under Contract No. DE-AC02-05CH11231. Portions of this research were carried out at the Stanford Synchrotron Radiation Lightsource, a national user facility operated by Stanford University on behalf of the U.S. Department of Energy, Office of Basic Energy Sciences. Personnel expenses were supported in part by the National Science Foundation Graduate Fellowship.

**Supporting Information**
Experimental details, details of theoretical calculations, TEM, SEM and SAED images. This material is available free of charge via the Internet at http://pubs.acs.org.